# On the energy of vanishing flow for colliding nuclei leading to same compound nucleus.

Supriya Goyal and Rajeev K. Puri*
*Department of Physics, Panjab University, Chandigarh-160014, INDIA*
*\* email: rkpuri@pu.ac.in*

## Introduction

System size effects in reaction dynamics is one of the most interesting topic in heavy-ion collisions. The effect of system size on the reaction dynamics from low to ultra-relativistic energies has been largely reported in the literature [1]. Among various observables and non-observables in heavy-ion collisions at intermediate energies, collective transverse flow is one of the most sensitive and sought after phenomena. This is also known as *in-plane flow*. At low incident energies, attractive interactions dominate and as the incident energy increases, repulsive interactions take over. As one goes from low incident energy to higher one, there exists a particular energy at which the net in-plane flow disappears. This energy is termed as the energy of vanishing flow (EVF) [2]. The dynamics of heavy-ion reactions depends on the asymmetry of the reaction [3]. The dependence of the EVF on the system size for nearly symmetric systems has been investigated extensively during the last few years and is found to depend strongly on the composite mass of the system by power law scaling ($\propto A_{tot}^{\tau}$) [4]. But the effect of mass asymmetry on the system size dependence of EVF is not studied anywhere in the literature. It is also known that the asymmetric reactions do not follow the same power law behaviour as followed by the symmetric reactions in the first place. Therefore, we plan to study the system size dependence of EVF for colliding nuclei with varying asymmetry. We plan to address this question using Quantum Molecular Dynamics (QMD) model [5].

## Model

The QMD model simulates the heavy-ion reactions on event by event basis. This is based on a molecular dynamic picture where nucleons interact via two and three-body interactions. The nucleons propagate according to the classical equations of motion:

$$\frac{d\mathbf{r}_i}{dt} = \frac{dH}{d\mathbf{p}_i} \text{ and } \frac{d\mathbf{p}_i}{dt} = -\frac{dH}{d\mathbf{r}_i}, \quad (1)$$

where H stands for the Hamiltonian which is given by

$$H = \sum_i \frac{\mathbf{p}_i^2}{2m_i} + V^{tot}. \quad (2)$$

Our total interaction potential $V^{tot}$ reads as

$$V^{tot} = V^{Loc} + V^{Yuk} + V^{Coul} + V^{MDI}, \quad (3)$$

where $V^{Loc}$, $V^{Yuk}$, $V^{Coul}$, and $V^{MDI}$ are, respectively, the local (two and three-body) Skyrme, Yukawa, Coulomb and momentum dependent potentials. The EVF is calculated by using the quantity "*directed transverse momentum* $<P^{dir}_x>$, which is defined as:

$$\left\langle p_x^{dir} \right\rangle = \frac{1}{A} \sum_{i=1}^{A} sign\{y(i)\} p_x(i). \quad (4)$$

Here $y(i)$ is the rapidity and $p_x(i)$ is the transverse momentum of $i^{th}$ particle. The rapidity is defined as:

$$y(i) = \frac{1}{2} \ln \frac{E(i) + p_z(i)}{E(i) - p_z(i)}, \quad (5)$$

where $E(i)$ and $p_z(i)$ are, respectively, the total energy and longitudinal momentum of $i^{th}$ particle. The EVF was then deduced using a straight line interpolation.

## Results and discussion

The asymmetry of a reaction is defined by the parameter called asymmetry parameter ($\eta$) and is given by:

$$\eta = \left|\frac{A_T - A_P}{A_T + A_P}\right|, \quad (6)$$

where $A_T$ and $A_P$ are the masses of target and projectile, respectively. The $\eta = 0$ corresponds to the symmetric reactions and nonzero values of $\eta$ defines different asymmetries of a reaction. We simulated various reactions for 1000-5000 events in the incident energy range between 90 to 350 MeV/nucleon. Keeping the total mass of the system fixed as 40, 80, 160, and 240, the value of $\eta$ is varied from 0 to 0.7. Calculations are performed at a reduced impact parameter ($b/b_{max}$) of 0.25. For the present study, we employed a soft equation of state (K=200 MeV) with momentum dependent interactions along with energy dependent cugnon cross-section. In fig. 1, we display EVF as a function of combined mass of the system for $\eta$ = 0.1, 0.3, 0.5, and 0.7. Different symbols are explained in the caption of the figure. Lines are power law fits ($\propto A_{tot}^{\tau}$). It is clear from the figure that all values of $\eta$ follow the power law behaviour. The values of $\tau$ are -0.3019, -0.2993, -0.2829, and -0.2719, respectively, for $\eta$ = 0.1, 0.3, 0.5, and 0.7. The value of $\tau$ is nearly same for every $\eta$. Our studies has shown that the mass dependence of EVF show mass asymmetry independent behaviour. This is due the fact that with increase in asymmetry, the number of nucleon-nucleon collisions and repulsive Coulomb interactions decreases, therefore EVF increases. The increase in EVF with asymmetry is more for lighter nuclei as compared to heavier nuclei due to the further less magnitude of Coulomb repulsions in lighter nuclei as compared to heavier ones.

## Acknowledgments

This work is supported by grant from Department of Science and Technology (DST), Govt. of India.

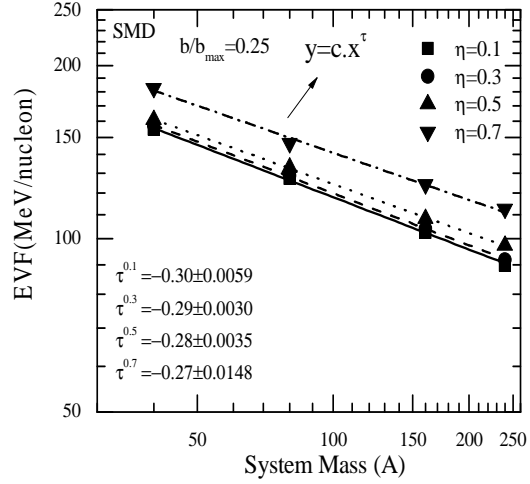

**Fig. 1** The EVF as a function of combined mass of system. The results for different asymmetry parameters $\eta$ = 0.1, 0.3, 0.5, and 0.7 are represented, respectively, by the solid squares, circles, triangles, and inverted triangles. Lines are power law fit ($\propto A_{tot}^{\tau}$).

**References:**

[1] I. Dutt and R. K. Puri, Phys. Rev. C **81**, 044615 (2010); I. Dutt and R. K. Puri, Phys. Rev. C **81**, 047601 (2010); I. Dutt and R. K. Puri, Phys. Rev. C **81**, 064608 (2010); I. Dutt and R. K. Puri, Phys. Rev. C **81**, 064609 (2010); D. Sisan *et al.*, **63**, 027602 (2001); Y. K. Vermani, S. Goyal, and R. K. Puri, Phys. Rev. C **79,** 064613 (2009); Y. K. Vermani, J. K. Dhawan, S. Goyal, R. K. Puri, and J. Aichelin, J. Phys. G **37**, 015105 (2010).

[2] D. Krofcheck *et al.*, Phys. Rev. Lett. **63**, 2028 (1989); H. Stöcker and W. Greiner, Phys. Rep. **137**, 277 (1986); G. D. Westfall *et al.*, Phys. Rev. Lett. **71**, 1986 (1993).

[3] B. Jakobsson *et al.*, Nucl. Phys. A **509**, 195 (1990); H. W. Barz *et al.*, Nucl. Phys. A **548**, 427 (1992).

[4] A. D. Sood and R. K. Puri, Phys. Rev. C **79,** 064618 (2009) and references therein.

[5] J. Aichelin, Phys. Rep. **202,** 233 (1991).